# Developing the Statistical Mechanics of Evolving Assembly Spaces

Anisha K. Yeddanapudi[1†], Abhishek Sharma[1†], Ian Seet[1], and Leroy Cronin[1*]

[1]School of Chemistry, The University of Glasgow, Glasgow, G12 8QQ, UK

*Corresponding author email: Lee.Cronin@glasgow.ac.uk †Equal contribution.

**Abstract**

A gap exists in connecting biological evolution within the framework of statistical mechanics. It has been hoped that statistical mechanics would provide us with a general predictive principle for the emergence of novelty and complexity. However, the emergence of novel functional features initiated by stochastic processes remains difficult to describe, because these processes commonly arise outside near-equilibrium or linear-response regimes. To understand how environmental influences generate diverse, open-ended, complex objects by driving physical systems to specific regions of phase space over long spatial and temporal scales, a new approach to non-equilibrium thermodynamics is necessary. Here, we present the coupling of assembly theory (AT) with statistical mechanics as a framework that works within the laws of thermodynamics, but can contextualise the processes of novelty, complexity, and evolution within these laws. We map construction processes from physical space to assembly space (AS) and introduce the concept of an assembler, a machine that constructs objects contingent on previous states and is represented by bitwise operations. Using this approach, we apply principles from stochastic thermodynamics with AT to define selection within large combinatorial spaces[1] and estimate the work required to create novel and complex objects. This approach allows us to represent complex processes, such as selection, within large combinatorial spaces as a set of rules for a memory-feedback loop followed by the assembler. This framework concretely defines the relationships between thermodynamics and information theory by introducing a novel aspect of physics that uses history and causal contingency to influence the probabilistic time evolution of complex systems.

## Introduction

Natural selection provides a foundational framework for understanding the differential existence of some organisms over others within evolving populations[2]. The fundamental laws of physics are usually formulated as fixed rules and describe how systems evolve from given initial conditions under those rules. While statistical mechanics, in principle, can predict the evolution of a system towards equilibrium, it offers limited insight into the spontaneous emergence, symmetry breaking, and persistence of biological information where effective rules are history-dependent, such as replication and mutation as described in the quasispecies model[3–7]. The emergence of novel functional features[8] – such as an organism's phenotypic traits – remains unexplained within the current framework of physical laws, which provides limited mechanistic insight into processes leading to the emergence of complexity and adaptation. Although genotypic effects that lead to functional phenotypic traits – such as point mutations – can be discussed using statistical mechanics, these processes arise outside near-equilibrium or linear-response regimes[9,10]. In equilibrium statistical mechanics, the number of accessible microstates, or equivalent phase space volume for given macroscopic constraints, grows exponentially as a function of the degrees of freedom, leading to the Boltzmann-Gibbs entropy functional; $S = k_B \ln\Omega$. However, biological or other far-from-equilibrium systems with long-range coevolving interactions occupy a small subset of their phase space through energy flux and low-entropy configurations and, hence, do not exhibit this exponential phase space growth[11,12]. Therefore, while equilibrium processes could, in theory, generate complexity, the necessary timescales would be infeasible, as the system would need to traverse its entire super-exponential phase space[13,14]. Externally driven non-equilibrium systems exhibit a hierarchy of coupled timescales, for example charge dynamics in electrochemical systems consist of multiple length and time scales, including Debye relaxation, diffusion, and electrochemical kinetics timescales. Hence, determining the characteristic time, length, and energy scales that govern the dynamics of complex systems subject to strong external drives or variable boundaries remains a non-trivial problem. Whether it be nanoparticle arrays at solid interfaces[15] or networks of protein filaments[16] it has long been recognized that a N-body system can exhibit phenomena spanning many orders of magnitude in both length and time[17,18].

Nevertheless, there has been a great interest in using non-equilibrium thermodynamics to establish fundamental principles for dynamical systems[19–22]. Some approaches, such as the theory of the adjacent possible[23] aim to address the challenges posed by highly combinatorial phase spaces. However, these approaches typically do not account for the effects of selection within the system due to either the system itself or environmental factors. Approaches such as variational principles on these systems, such as Maximum Caliber (MaxCal)[24–28], rarely work for complex biological processes where the dynamical constraints are unknown. In conjunction with MaxCal, the extension of near-equilibrium concepts to far-from-equilibrium has been largely enabled by fluctuation theorems (FTs), an application of large deviation theory to non-equilibrium systems[29]. The most notable FTs are Crooks' FT[30] and Jarzynski's inequality[31], which have seen extensive experimental verifications, particularly in solidifying the relationship between thermodynamics and information theory[32–34]. These FTs have proven effective for small systems, but applying them to large non-Markovian processes, especially those influenced by many unknown external control parameters, remains a major challenge. Building on these FTs, there has been significant progress in tailoring these thermodynamic frameworks to biological processes. Recent works, particularly those of England and Kolchinsky, have uncovered fundamental bounds on the thermodynamic costs of several biological processes such as selection[35], adaptation[36,37] and replication[38–40]. However, these works have yet to investigate the thermodynamics of physical systems that produce novelty and their application to general biological systems remains limited.

Consider, for example, any non-Markovian biological or complex process such as sensory adaptation. If we proceed to apply any of the currently known FTs to this system (which includes the approaches of England and Kolchinsky), the memory-dependent nature of these processes will make it incredibly difficult to calculate the probability distributions of any forward trajectory, $P(w)$. Additionally, when these systems operate outside the linear-response regimes, the external constraints, represented as Lagrange multipliers in the MaxCal framework, are no longer clearly defined, preventing us from deriving any meaningful distribution using this approach. Even if we can numerically approximate the distribution of a forward trajectory, determining the time-reversed probabilities $P(-w)$ remains a separate challenge. Retracing the exact contingent processes that led the system to its current state is nearly impossible for non-Markovian dynamics, since the time-reversed path is not simply the reversal of the forward one and depends on the system's full history.

Here, we introduce Assembly Theory (AT)[41] within a statistical mechanics framework to extend stochastic thermodynamic principles to non-Markovian processes with variable length, time, and energy scales. In the context of AT, an object is defined as a finite, distinguishable entity that persists over time such that it can be observed and decomposed to infer its construction pathway. The assembly space (AS) for an object begins with a set of elementary building blocks which are recursively joined to form intermediate structures which can be used for subsequent construction steps. Each object, $i$, is characterized by an assembly pathway, and assembly index, $a_i$ which denotes the minimum number of steps required to generate the object from its building blocks. We characterize objects in AT using their shortest path because of the computational intractability of the most probable path due to unknown environmental constraints over vast spatiotemporal regions. For this reason, the shortest path remains the most unbiased calculable representation of an object, which makes it an intrinsic property of the object given fixed building blocks and joining operations. For a given observable volume, we define copy number as the number of exact copies of an object. Similarly to the single object, when multiple objects are observed, the assembly space, called Joint Assembly Space (JAS) is defined by the shortest path needed to create all the objects simultaneously. By combining both assembly index and copy number, the amount of causation required to construct the observed ensemble can be quantified, which is complemented by the assembly equation (1) and represents the total amount of selection necessary to produce an observed ensemble.

$$A = \frac{\Omega}{N_T} \sum_i e^{a_i}(n_i - 1) \tag{1}$$

where, $a_i$ is the assembly index of $i^{th}$ object, $n_i$ is the copy number, $N_T$ represents the total number of objects in the ensemble, and Ω is the constant with units of kJ/mol.

We start with mapping the observables in physical space to assembly space. Within the assembly space, we define a new framework for a generalized relationship between complexity, time and information. When a non-equilibrium, exploratory system is mapped into AS, we enable the use of variational principles that would otherwise be inaccessible. The motivation of this approach arises because of the highly combinatorial expansion of complex systems. These combinatorial spaces can be tamed in AS, due to the linear progression of time as well as contingent processes being ingrained within the AS representation of any physical system[42]. From here, we can apply our framework to study the thermodynamic properties of open-ended systems undergoing selection, by bounding the amount of work required to process the information needed for a computational thermodynamic construction process to create novel objects in AS. Finally, we discuss the relevance of this approach in studying the physics of patchy particle simulations under external drive.

## Mapping between Physical Space and Assembly Space

To better describe the process of mapping between physical and assembly space, let us simplify a general forward dynamic process leading to the formation of complex objects with a representative chemical reaction network (CRN). The key features of CRN include a finite set of chemical species (reagents and products), a list of reactions representing how these species are transformed, and kinetic parameters that quantify the rates of the various reactions. Consider a CRN with a finite set of reactive species $\boldsymbol{\mathcal{S}} = \boldsymbol{X_1}, \boldsymbol{X_2}, \ldots, \boldsymbol{X_n}$ and finite set of reactions $\boldsymbol{\mathcal{R}} = \boldsymbol{R_1}, \boldsymbol{R_2}, \ldots, \boldsymbol{R_r}$. The general equation for a CRN for $n = |\boldsymbol{\mathcal{S}}|$ reactive species is given by,

$$\frac{d\mathbf{c}(t)}{dt} = \mathbf{S}\mathbf{v}\big(\mathbf{c}(t)\big) \tag{2}$$

where, $\mathbf{c}(t) \in \mathbb{R}^{\boldsymbol{n}}_{\geq \mathbf{0}}$ is a vector of the concentrations of various species such that $\mathbf{c}(t) = \big(\mathbf{c_1}(t), \mathbf{c_2}(t), \ldots \mathbf{c_n}(t)\big)$, $\mathbf{S} \in \mathbb{R}^{\boldsymbol{n} \times \boldsymbol{r}}$ is the stoichiometric matrix and $\mathbf{v}(\mathbf{c}) \in \mathbb{R}^{\boldsymbol{r}}_{\geq 0}$ is the vector of reaction rate functions. A single snapshot of the state of CRN at time $t = t_0$ at which measurement is performed is defined by $\boldsymbol{\mathcal{S}}$ and $\mathbf{c}(t_0)$. Often for complex systems such as large biochemical networks, their behaviour is defined by replication-mutation dynamics without the knowledge of complete mechanistic insight, $\boldsymbol{\mathcal{S}}$ is not fixed, and the reaction rates are highly dynamic and evolve with time. The potential space of species that can form is highly combinatorial. AT tracks the forward process in physical space either as a single snapshot in time or over a period of time, by mapping the features of observed objects and their copy number into assembly space. Consider a simple CRN with a fixed number of species for simplicity, $\boldsymbol{\mathcal{S}}$ represents the list of measured objects which are represented by their assembly pathways (with $\boldsymbol{a} = (a_1, a_2, \ldots, a_n)$ as their assembly indices) and $\mathbf{c}(t_0)$ is the list of copy numbers of the observed objects. Assembly pathways are the minimal pathways to construct an observed object or an ensemble of objects by recursive operations independent of the physical mechanism that constructed them. The mapping between physical space and assembly space as an outcome of measurement is shown in Figure 1A. To quantify the amount of causation associated with an observable, by mapping physical space to assembly space, the construction of the observed objects can be represented by a finite set of discrete recursive operations. These construction operations within a large combinatorial space can be defined by a thermodynamic state-machine engine called an assembler performing work by dissipating heat to reach the observed ensemble of objects[43], see Figure 2B.

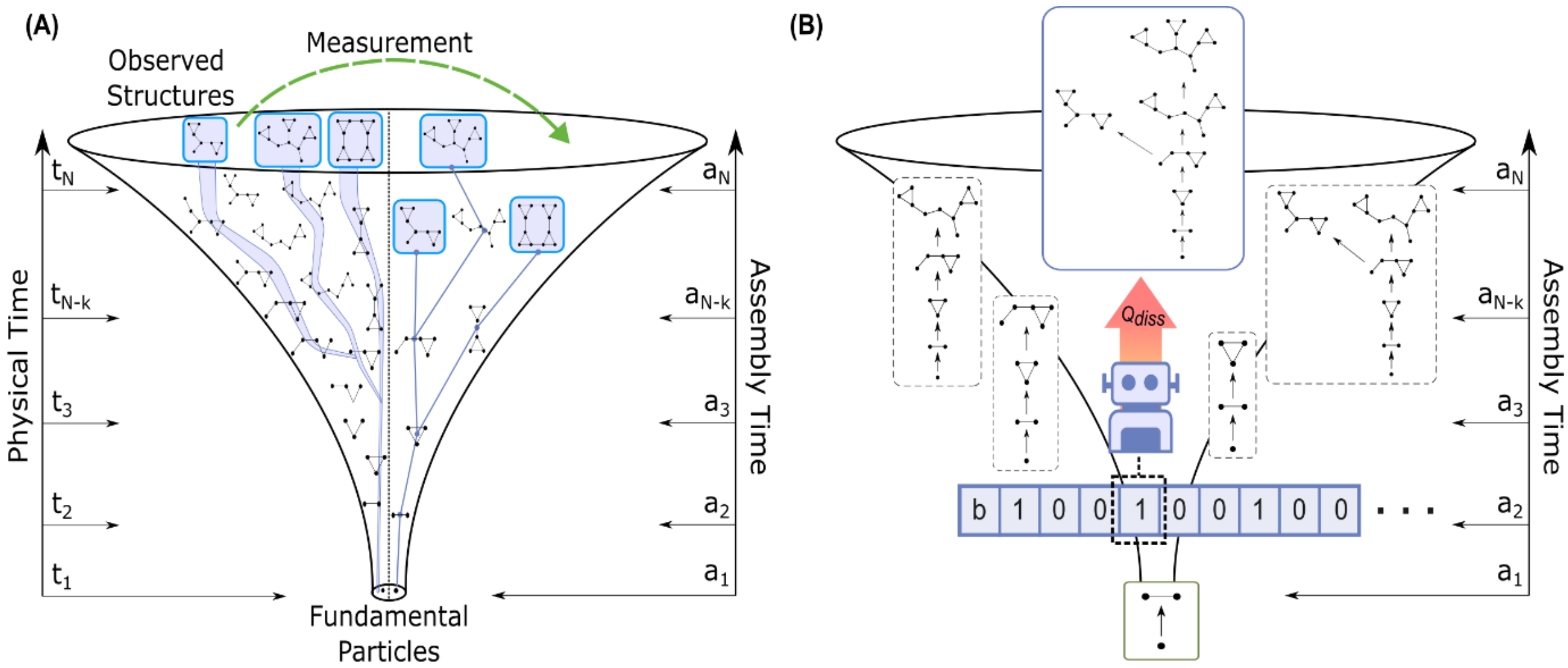

**Figure 1: Physical to Assembly Space Mappings and the Assembler** **(A)** Pictorial depiction of the physical space and its equivalent assembly representation. The process of measurement coupled with a reverse computational process (see text for details) allows us to move between physical and assembly spaces. Complex physical processes may traverse multiple unknown pathways to reach the observed objects; however, the assembly representation remains unbiased by choosing the shortest path. **(B)** Using the matrix representation, a Finite State Machine (FSM) can traverse the assembly space by applying bit-wise operations to reach a desired JAS or AS. As the machine moves towards the final state, it dissipates heat, allowing us to calculate the energetics of this process (see text for details).

**Equilibrium Ensembles in Assembly Space**

As discussed in the previous section, at a given time, mapping a forward process to the assembly space (AS) gives the list of unique objects $\boldsymbol{\mathcal{S}}$, their assembly indices $\boldsymbol{a}$, and copy numbers $\boldsymbol{c}$. With these, the assembly state of an ensemble at a given time is defined as $\boldsymbol{\mathcal{L}} = (a_1, c_1), (a_2, c_2), \dots, (a_n, c_n)$ for $n$ unique objects. A key characteristic of ASs is that they are super-exponentially growing combinatorial spaces, hence, more objects exist in assembly space than can be built in finite time with provided resources. Therefore, historical contingency is an intrinsic component of the space. To construct the AS of an observed structure, it is recursively broken down into its building blocks. Since we do not typically know the full assembly trajectory of an object, we instead find the minimal number of steps to produce the object using recursive joining operations starting from the building blocks. Additionally, every object generated in the process of construction belongs to an assembly pool and can be used at any time along the assembly path. During the construction of ASs and JASs for an observed structure, there will be several objects present in the assembly path that are not physically observed. For example, consider the assembly space of a linear string of length 6; $1 \xrightarrow{1} 2 \xrightarrow{1} 3 \xrightarrow{3} 6$. In this pathway, strings of length 1, 2, and 3 were not observed in the physical system; however, they form the contingent pathway for constructing 6. These intermediate structures are defined as virtual objects within the assembly space.

To apply a statistical physics framework to AT, we must define a mathematical representation for the generalised assembly pathway. Similar approaches are used for networks, where graphs are represented by adjacency matrices and constraints are applied to those matrix representations[44]. Similarly, we can describe the AS (independent of copy number) as the adjacency matrix of a digraph, denoted as $L$, with a predefined ordering in the rows and columns. Additionally, we can also generate the JASs and ASs using this matrix representation. Consider the ASs and matrix representations of all possible non-isomorphic graphs with $N = 3$ nodes shown in Figure 2. To introduce copy number, we incorporate the diagonal matrix $C$ defined in equation 2, such that $c_i$ is the copy number of object $i$ . Therefore, the full assembly matrix of a particular object's AS or JAS is denoted as $L + C$ such that $C_{ij}$ can be defined as,

$$C_{ij} = \begin{cases} 0 \text{ else} \\ c_i \text{ if } i = j \end{cases} \tag{2}$$

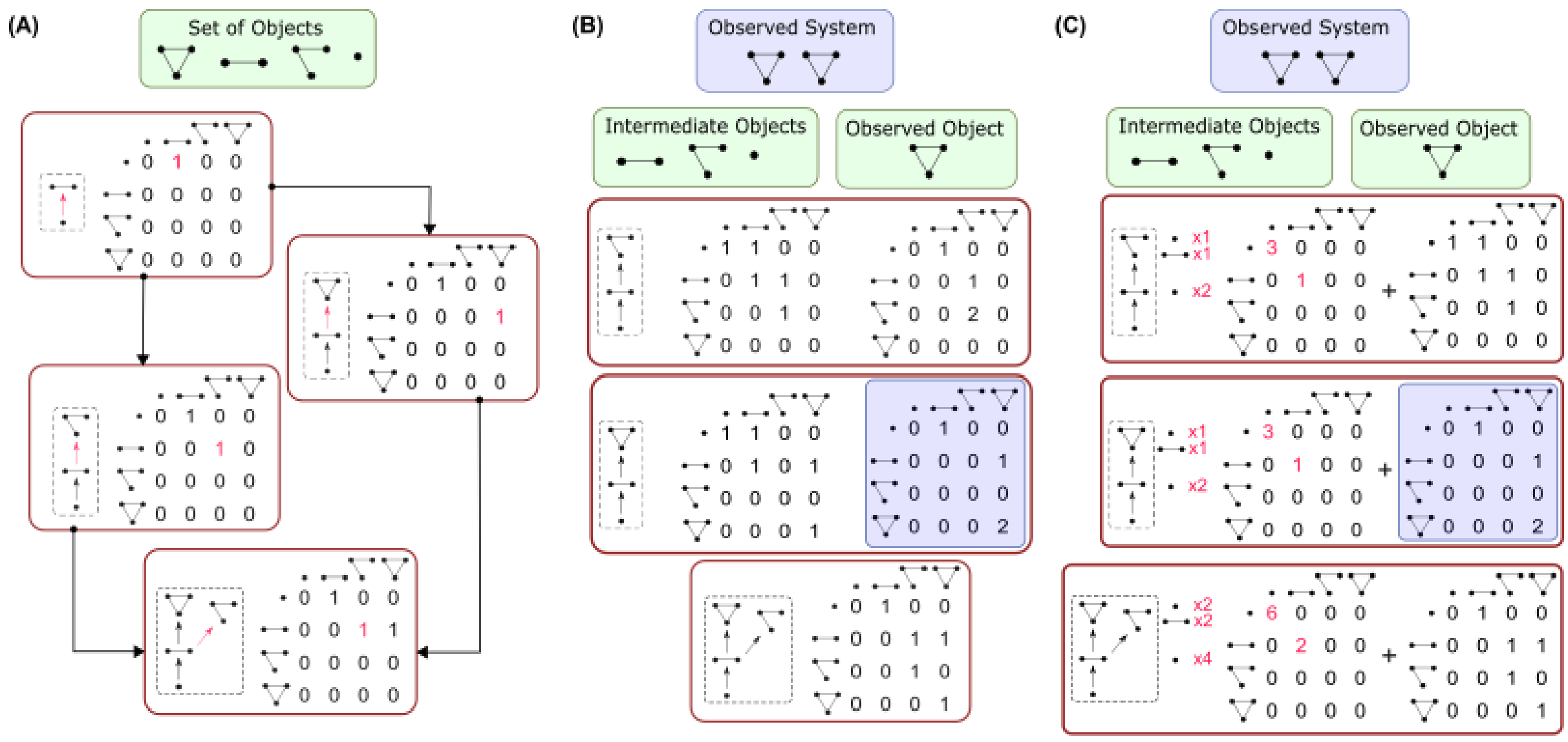

**Figure 2: Matrix Representation & Bitwise Evolution of Microstates in Assembly Space** **(A)** Depiction of how each JAS or AS given a complete set of objects can be represented as matrices, where the objects are ordered by some metric (maximum length for linear strings or graph energy). Additionally, we also show how a FSM would traverse a set of ASs/JASs by using bit-wise operations on the upper triangular portion of the L matrix. **(B)** We can also introduce copy number on the diagonals given some resource constraints (particle number) based on an observed state. Doing so also allows us to highlight a desired assembly matrix state (in blue) as a representation of the physical system. The goal of our FSM is to proceed with transitions such that we can consistently reach this final state (see text for details). **(C)** The copy number of virtual objects (virtual copy number) used in the construction process of the final ensemble can be considered as weights added to the copy number diagonal. The grand canonical ensemble can also be studied under the influence of virtual copy number (see SI Sec 1.2 and Fig. S2 for details).

A critical aspect of this matrix representation is how it allows us to construct a generalized AS by simply applying bit-wise operations, given construction rules. Suppose we want to construct an AS or JAS given some resource bounds. These bounds would be present along the diagonals by limiting the copy number of specific objects. The bit-wise trajectory for a finite state machine (FSM), which acts as a model of a physical process that dissipates heat to construct a JAS for all graphs with $N = 3$ nodes is given in Figure 2. Therefore, a generalized ensemble of ASs can be traversed by a Turing machine via operations on the matrix. The matrix representation and the ability for an FSM to traverse the assembly state space allows us to apply MaxEnt and other statistical mechanical principles to objects in AS[45,46]. This methodology follows closely with approaches applied to study the statistical mechanics of networks[44]. For example, fixing the expected number of edges from constraints applied to the adjacency matrix of a network is equivalent to fixing the expected assembly in our framework. We will discuss these principles in greater detail in the following section.

**MaxEnt Principle & Equilibrium Distributions in Assembly State Space**

We know from equilibrium statistical mechanics that by applying the maximum entropy principle (MaxEnt) under specific constraints, we can derive the Boltzmann distribution[47]. Using the matrix formulation above and principles commonly applied in the statistical mechanics of networks, we can derive equivalent equilibrium distributions of systems in the assembly space[44].

Let $M \in \mathcal{A}$ be an AS or JAS represented as a digraph in our set of assembly spaces, $\mathcal{A}$. Let *P(M)* be the probability of graph *M* within our ensemble. Following from statistical physics, the best choice of probability distribution is one that maximizes the Gibbs entropy:

$$\mathrm{S} = -\sum_{\mathrm{M} \in \mathcal{A}} \mathrm{P(M)} \ln P(M) \tag{3}$$

subject to constraints

$$\langle x_i \rangle = \sum_{M} P(M) x_i(M) \tag{4}$$

along with the normalisation condition.

We can proceed by maximising the following functional with the introduction of Lagrange multipliers $\nu, \lambda_i$, performing the functional derivative, and equating it to zero.

$$\mathcal{L} = S + \sum_i \lambda_i \left( \langle x_i \rangle - \sum_M P(M) x_i(M) \right) + \nu \left( \sum_M P(M) - 1 \right) \tag{5}$$

This gives

$$P(M) = \frac{e^{-H(M)}}{Z} \tag{6}$$

where *H(M)* is the assembly Hamiltonian

$$H(M) = \sum_i \lambda_i x_i(M) \tag{7}$$

and *Z* is the partition function

$$Z = e^{\nu+1} = \sum_{\mathcal{A}} e^{-H(M)} \tag{8}$$

Eqns 5-7 define the distribution over a specified set of ASs that maximizes the entropy subject to known constraints, this can form as the null or baseline ensemble where selection is detected by deviations from this generic accessible space. This is an exact analogue of the Boltzmann distribution for ASs. Using the model above, the expected value of any property *x* in AS is

$$\langle x \rangle = \sum_{\mathcal{A}} P(M) x(M) \tag{9}$$

Usually, average properties of the assembly ensemble can be derived by taking appropriate derivatives of the analytical formulation of the partition function. However, obtaining this result can be a challenge, and this is the case for the grand canonical ensemble. Therefore, numerical approaches can be used to approximate specific properties of the ensemble.

An intrinsic property of our assembly ensemble is the assembly index, which is calculated by summing the upper triangular entries of our matrix. Consider that we know only the variance of the average assembly index across our ensemble. This denotes an assembly representation of the canonical ensemble in statistical physics. In this case our assembly Hamiltonian would take the form

$$H(M) = \lambda\left(\sum_M P(M)(a(M) - \bar{a})^2 - \sigma^2\right) \tag{10}$$

Here $\bar{a}$ is the mean assembly of the ensemble which is a constant and has the full representation

$$\bar{a} = \frac{\sum_M a(M)}{\text{card}\ (\mathcal{A})} \tag{11}$$

where card $(\mathcal{A})$ represents the total number of graphs/matrices in the ensemble (cardinality of our state space $\mathcal{A}$). We then have the following probability distribution.

$$P(M) = \frac{1}{Z} e^{-\lambda(a(M)-\bar{a})^2} \tag{12}$$

$$Z = \sum_M e^{-\lambda(a(M)-\bar{a})^2} \tag{13}$$

In a physical system limited by $\bar{a}, \sigma^2$, the complexity and exploration of the space is confined. Specifically, by constraining the average assembly of the microstates, we limit the complexity of the objects. Constraining the variance confines the exploratory ability of the state space. Therefore $\bar{a}$ and $\sigma^2$ confine the length and breadth of possible JASs, and each works as a trade-off to the other. For JASs with long lengths, which represent the right end of the Gaussian distribution in **Error! Reference source not found.**, the current object is joined with itself in a linear growth process creating the most complex structure, i.e. $1 \rightarrow 2 \rightarrow 4 \rightarrow 8$. On the other hand, we could have bifurcating processes where the JASs grows by exploring the entire space at a lower complexity (as defined by assembly index).

In the canonical case, we can obtain an analytical representation of the partition function which allows us to formalize the physical representation of $\lambda$. For the calculation below, note that for the matrix *M* the summation of the upper triangular portion is, $a(M) = \sum_i m_{i,i+1}$where $m_{ij}$ are the indexed elements of matrix *M*.

$$Z = \sum_M e^{-\lambda(a(G)-\bar{a})^2} = \left(e^{-\lambda\bar{a}^2} + e^{-\lambda\left(3-2\bar{a}+\bar{a}^2\right)}\right)^{\text{card}(\mathcal{A})} \tag{14}$$

From our definition of λ in Sec λ & Characteristic Informational Work, we use the Lagrange multiplier $\kappa = \lambda^{-1}$ instead of λ to ensure that for long linear strings *N*, $\kappa \to -\beta$, similar to the Boltzmann distribution.

Now, the formulation above quantifies an equilibrium probability distribution across all possible ASs and JASs given an observed physical system, independent of copy number. To introduce copy number, we must move to the grand canonical ensemble.

Now, let us consider the case where we constrain the average concentration $\langle N_i \rangle$ of each structure (object) $i$ in the assembly space. These define soft constraints on the assembly space matrix, where $c_i(M)$ is the copy number of structure $i$ for assembly matrix $M$. Then, the Hamiltonian takes the form

$$H(M) = -\lambda(a(M) - \bar{a})^2 - \sum_i \mu_i c_i(M) \tag{15}$$

Performing the functional derivative and setting it equal to zero, we obtain a new probability distribution:

$$P(M) = \frac{1}{Z} e^{-\kappa(a(M)-\bar{a})^2 - \sum_i \mu_i c_i(M)} \tag{16}$$

$$Z_{GC} = \sum_M e^{-\kappa(a(M)-\bar{a})^2 - \sum_i \mu_i c_i(M)} \tag{17}$$

Due to the nature of the Lagrange multipliers $\mu_i$ it becomes difficult to obtain a generalized analytical expression for $Z_{GC}$, so we approximate $Z_{GC}$ numerically.

Additionally, the Lagrange multipliers $\mu_i$ are used to match our distribution to a known evolving physical system. In such a case, we can alter the $\mu_i$values such that the distribution matches the copy number evolution of various structures in the physical system. However, for a general study of the distributions, we set $\mu_i = 1$ for all $i \in \mathcal{A}$, since we are not matching to a specific physical system. Additionally, in the case of emergent phenomena the system is dynamically evolving. For example, consider an open-ended chemical reaction network, it is almost impossible to know beforehand exactly what reactions will happen when, therefore determining the exact values of all these Lagrange multipliers is not possible. For this reason, we consider approximations based on general information about the physical system and present an overall theoretical formalism.

We analysed the distribution as a function of particle number and temperature, which is shown in **Error! Reference source not found.**. Temperature in this analysis represents an effective Lagrange multiplier which acts as a control parameter for the ensemble, similar to its definition in the thermodynamics of computation[48]. We see that the average assembly shifts as a function of increasing maximum chain length, reflecting how larger linear strings are, generally, more complex than shorter ones. Additionally, this provides more numerical evidence that the assembly as a function of string length is $a \sim \log_2 N$. We note that as the temperature increases, the probability equalises across the state space. Therefore, as more energy is added into the system, physical barriers caused by complexity and copy number which exist at lower energies are relaxed. At higher temperatures, the system becomes independent of any physical restrictions, a common phenomenon exhibited in many chemical processes. The same relationship is also exhibited in the grand canonical ensemble, and can be seen in Sec 1.1 and Fig. S1.

The string assembly algorithm used in this paper is related to the graph assembly algorithm described in Seet et al.[49]Although it is possible for the graph assembly algorithm to operate on strings as a string can be expressed as a linear graph with node colourings, such an algorithm would be highly inefficient. Given that strings are a highly ordered data structure, it is logical to employ different heuristics and branch-and-bound strategies as compared to those used for graphs. The algorithm employs two major strategies to reduce space and time usage: the use of interval data structures and a string-specific branch-and-bound heuristic in which we use a precalculated rolling hash combined with a sliding window to construct a lower bound that is strictly less than the assembly index for a given string while being tighter than the bound described in Seet et al.[49]. Further information about the algorithm is available in the SI.

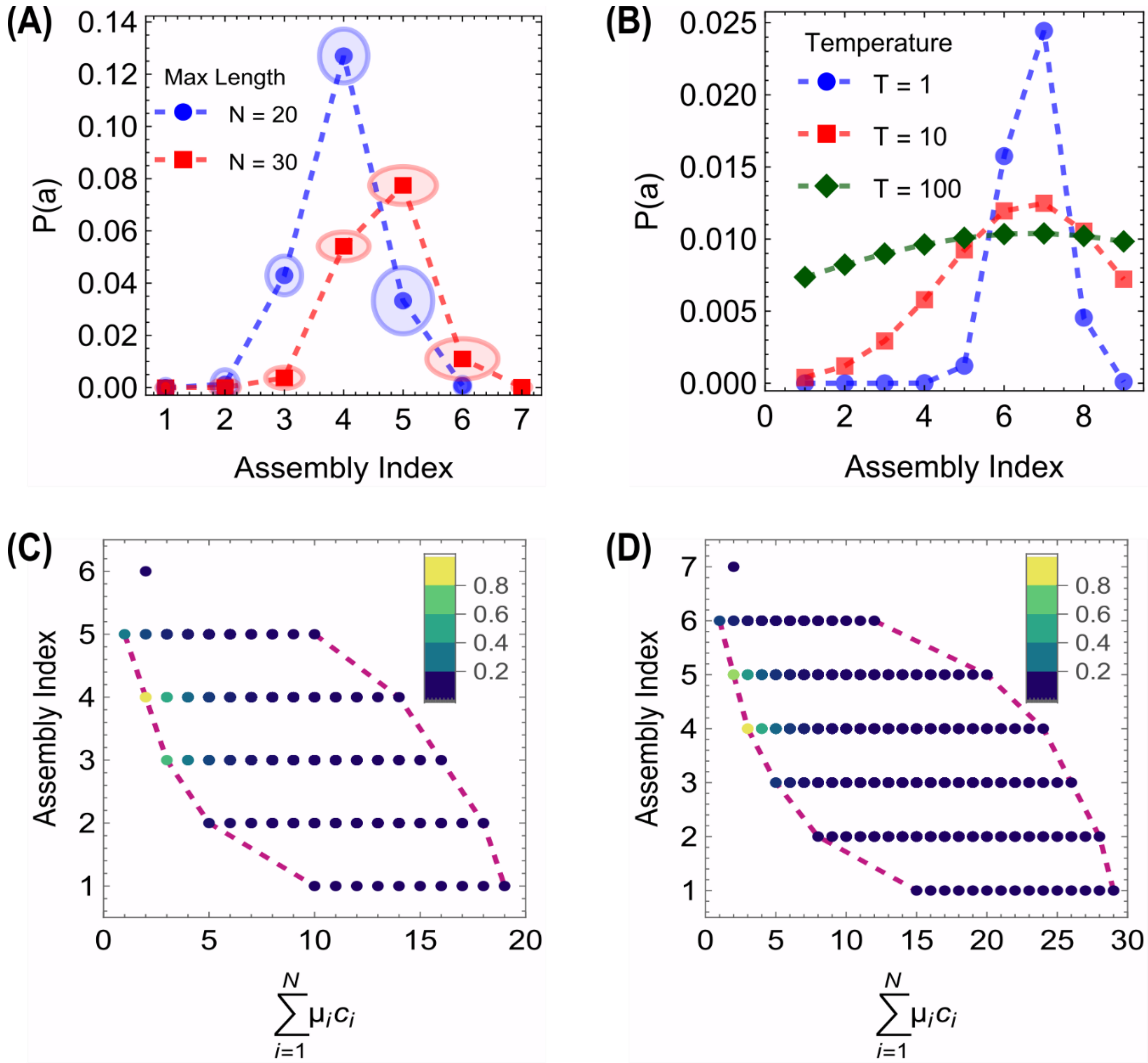

**Figure 3: Equilibrium Assembly Space Distributions (A)-(C)** The distributions of the canonical ensemble (Eqn 17) and grand canonical ensemble (Eqn. 22) under the influence of external parameters: temperature and maximum number of building blocks (maximum string length for canonical ensemble), N. **(A)** The canonical ensemble distribution as a function of maximum string length, N. The bubble size is proportional to the number of linear strings from length 1 to N which have a particular assembly index. **(B)** The canonical distribution for N = 100, as a function of increasing temperature. **(C)** The grand canonical distribution as a function of maximum string length, N = 20 and **(D)** N = 30.

**λ & Characteristic Informational Work**

Using the expression derived in Eqn 14, we can formulate a physical definition of our Lagrange multiplier, λ, as it relates to the characteristic informational work in assembly space. By using the matrix definition, moving forward in assembly space can be modelled as flipping values of $m_{ij} \in M$. Therefore, independent of copy number, changing the informational content (or history) of a structure is equivalent to flipping specified bits.

Bennett claimed that the work value of a string *S, WV(S)* is

$$WV(S) = \big(l(S) - K(S)\big)k_B T\, ln\, 2 \tag{18}$$

where $l(S) = N$ is the length of S and K(S) denotes the Kolmogorov complexity of S[50]. Now, considering the canonical ensemble, we can define the free energy as

$$F = \langle a \rangle - \lambda S = -\lambda\, ln\, Z \tag{19}$$

We know that the amount of work necessary to change the informational content of a linear string in assembly space is WV(S), so equating the change in free energy to work

$$\Delta F = -\lambda \cdot \mathrm{card}(\mathcal{A})\, ln\left(e^{-\lambda \bar{a}} + e^{-\lambda\left(3 - 2\bar{a} + \overline{a^2}\right)}\right) = WV(S) \tag{20}$$

We can solve for λ by assuming $\lambda \ll 1$ and Taylor expanding to obtain

$$\lambda = \frac{-k_B T\big(N - K(S)\big)}{\mathrm{card}(\mathcal{A})} \tag{21}$$

Let us consider the specific case of linear strings then $K(S) = log_2 N$ and $a(S) \sim log_2 N$ then Eqn 21 can be simplified to

$$\lambda = \frac{-k_B T(N - log_2 N)}{\mathrm{card}(\mathcal{A})} \tag{22}$$

We can also rewrite the above relation in terms of assembly, $a$ where $a(S) \sim log_2 N$ and $N \approx 2^a$.

$$\lambda = \frac{-k_B T(2^a - a)}{\text{card}(\mathcal{A})} \quad (23)$$

And if we approximate $\text{card}(\mathcal{A}) \sim cN$ where $c$ is a constant then in the large N limit, $\lambda \to -k_B T/c$.

However, equilibrium ensembles in assembly space do not fully describe the thermodynamic relationships that exist among resources, contingency, and causality. To approach this task, we will need to draw from the thermodynamics of information and describe our system computationally.

**MaxEnt Maximum Assembly Index vs Maximum String Length Bound**

Here we take all binary strings of maximum length $N = 10$ and repeat the same analysis to generate the max $N$ and max $a$ distribution. In general, binary strings are a valid surrogate for complex physical systems such as template-based copolymerization where the joining operations are well-defined with respect to the template. We can see the same relationship for increasing λ in binary strings that is present for increasing temperature in linear strings. A comparison between the two possible distributions is shown in Figure 4C. Notice that the $N = 10$ and $a = 8$ distributions, which both take the entire generated state space of strings, are the same. Deviations between these two types of distributions occur when the state space is limited in some way, $N = 9$ vs $a = 5$.

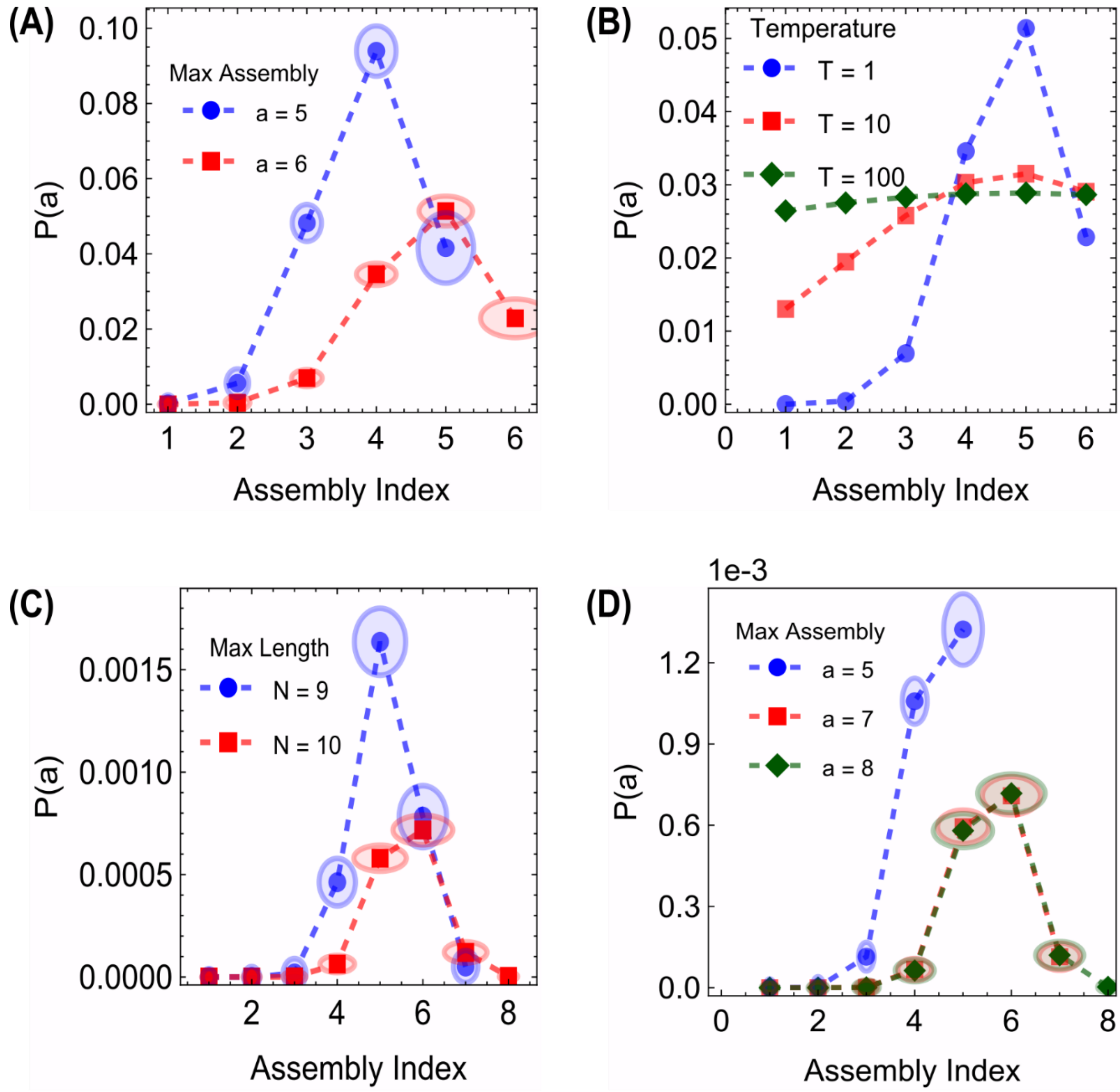

**Figure 4: Max Assembly vs Max String Length Distributions (A)** The probability distribution for all linear strings up to length, N = 100, considering a cut on maximum assembly index instead of a cut on maximum string length. The distribution for maximum string length is shown in **Error! Reference source not found.**A. **(B)** The temperature dependence of the probability distribution for linear strings given a maximum assembly cut. **(C)** The probability distribution for all binary strings of maximum length N = 10 with λ = 1 and T = 1 given the maximum length cuts of N = 9 and N = 10. **(D)** The distribution for the same binary string distribution with λ = 1 and T = 1 given various maximum assembly index cuts.

**Patchy Particle Simulations.**

Using the distributions derived in Sec. MaxEnt Principle & Equilibrium Distributions in Assembly State Space, we can describe the phase space collapse exhibited by a system under the influence of an external drive. The system we consider for this study is a 2D patchy particle simulation which consists of two charged central particles (A being positive and B being negative) with two attractive patches located at opposite poles. The central particles interact through a combination of Lennard-Jones and simplified Yukawa potentials, to model the electrostatic interactions, while the patches interact via a Morse potential, enabling bonding to form linear chains. To drive the system out of equilibrium, we introduce an oscillating external electric field in the positive x-direction. The positively and negatively charged particles interact with this field, leading to a collapse of the accessible phase space and the emergence of distinct nonequilibrium structures. For more details regarding the simulation refer to SI Sec 3.

The macroscopic organization of the particles depends strongly on the frequency of the external field, as shown in Fig 5A. At low frequencies, the particles arrange into alternating layers of opposite charge, referred to as Horizontal Stacking (HS). At high frequencies, the field oscillates too rapidly for the particles to orient themselves in ordered arrangements, resulting in a disordered state that resembles equilibrium, called Driven Random (DR). Between these two regimes lies a critical frequency range where the particles self-organize into charge-separated Vertical Stripes (VS). These macroscopic arrangements can be characterized by the KL-divergence between the non-equilibrium and equilibrium particle distributions, as shown in Fig 5D. The KL-divergence peaks at frequencies in the VS regime, indicating the largest difference from equilibrium, and decreases in the DR regime where the system most closely resembles equilibrium.

The macroscopic properties of the system directly influence the system's microscopic structures. This behaviour is captured by the partial structure factor shown in Fig 5B, which measures the excess probability of forming specific types of bonds. Since the system contains two particle species, A and B, three bond types are possible: A–A, B–B, and A–B. The charge segregation produced by the external drive increases the probabilities of bonds A-A and B-B over A-B. In equilibrium, all bond types remain equally probable.

The external drive shapes the system's ability to traverse its phase space. As discussed in Sec. MaxEnt Principle & Equilibrium Distributions in Assembly State Space, equilibrium systems tend to explore with high breadth and low length. Here, the system creates a larger number of unique structures at low assembly index and copy number, as shown in Fig 5C. Under nonequilibrium driving, however, the system follows a high length and low-breadth strategy. The macroscopic separation of the particle species favours a higher copy number of specific complex structures. This is shown in Fig 5C when comparing the equilibrium structures to those produced in the VS or HS regimes, in both cases, non-equilibrium driving results in the creation of long monotonic linear chains with high assembly and copy number. Additionally, a more detailed comparison between the various behaviours and chain lengths along with the radial distribution functions is provided in SI Sec 3 and Fig S3.

These shifts in exploration have direct consequences for the informational work required to realize the observed phase-space collapse. From Eqn 29, since equilibrium and non-equilibrium have the same cardinality of their ASs, we have $\lambda \propto 2^{a}$. Consequently, the larger assembly indices observed under nonequilibrium driving imply greater information-processing requirements. In other words, the externally driven system must perform more work to process the necessary information from its external environment to constrain its exploration and collapse its phase space.

Finally, using Eqn 1, we can calculate the assembly of our system and characterise the types of complexity that our system exhibits across its different regimes. As shown in Fig. 5E, assembly exhibits two prominent peaks: one associated with the VS regime and another within the DR regime. The HS regime generates comparatively low assembly despite producing structures with high copy numbers because these structures consist primarily of short monotonic chains, which possess relatively low assembly indices.

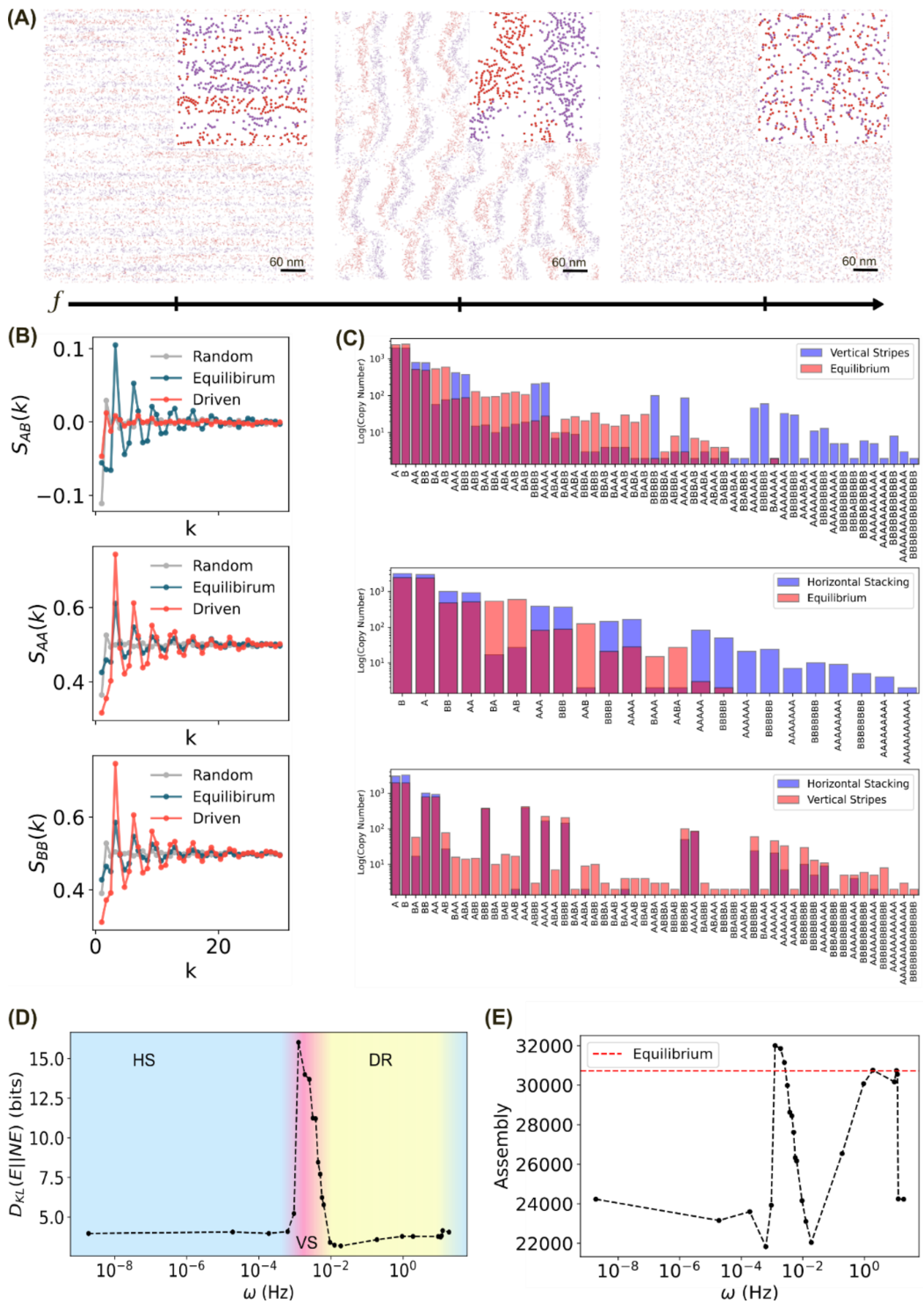

**Figure 5: Summary & Characterisation of Patchy Particle Simulations (A)** The 2D patchy particle simulations of 15,000 particles run for 1 million timesteps where the external drive is an oscillating E-field increasing in frequency over time, describing the formation of macroscopic ordering as a function of increasing frequency. A total of five independent seeded simulations were run (see SI for details). **(B)** The partial structure factors for each unit cell (A-A, A-B, B-B) within the intermediate linear chains for the critical frequency. **(C)** The structure-level distribution of the three different simulation types: equilibrium, constant electric field, and the critical frequency. **(D)** The KL divergence between equilibrium, denoted as E, and non-

equilibrium, denoted as NE, to characterise the macroscopic ordering as a function of frequency. The gradients depict the three types of macroscopic structures which emerge as a function of frequency, which are shown in (A); HS (Horizontal Stacking), VS (Vertical Stripes), and DR (Driven Random) **(E)** The assembly, shown in Eqn 1, of the ensemble as a function of frequency where the red line represents $A$ for the equilibrium distribution.

Since assembly has an exponential dependence on assembly index and a linear dependence on copy number, the lower copy number but higher assembly index strings generated in equilibrium outweigh those in HS. In contrast, the VS regime combines large copy numbers with long monotonic chains whose lengths drive assembly indices to much higher values, resulting in the largest assembly observed across all frequencies. The DR regime remains closest to equilibrium in terms of its microscopic structural properties, where the complexity is driven by the non-monotonicity of the strings.

These results demonstrate that the influence of an external environment can alter the macroscopic organization which fundamentally reshapes how the system explores its assembly state space. By shifting exploration to smaller subsets of favoured assembly spaces, the drive induces a phase-space collapse that increases the work required to process the necessary information to generate the observed structures in assembly space. In conclusion, nonequilibrium driving can function as a mechanism for generating complex organization through directed exploration of assembly space.

## Conclusions

We have introduced the statistical mechanics of AT and shown how it derives ensemble distributions of objects in AS. To do so, we discuss mappings from physical space to AS, built from two principles: the matrix representation of the assembly digraph coupled with copy number and an information-theoretic framework which connects the consequences of our AS distributions to information processing in physical space. We demonstrate how our framework allows a unified language for describing information theory in the context of the statistical mechanics of complex systems that process information from their external environment. The statistical mechanics of AT allows us to apply thermodynamic principles to systems where causal contingency through selection and phase space collapse starts to play a prominent role in their emergent properties. By incorporating this framework with stochastic thermodynamics and Markovian dynamics we can seek to develop new physics and mathematical principles to model and describe the time-series evolution of macroscopic observables for complex systems.

**Data Availability/Code Availability**

All the data required to produce the figures is available as a supplementary data file and a supplementary document explains how the figures were made

**Acknowledgements**

We acknowledge financial support from the John Templeton Foundation (grant nos. 61184 and 62231), the Engineering and Physical Sciences Research Council (EPSRC) (grant nos. EP/L023652/1, EP/R01308X/1, EP/S019472/1 and EP/P00153X/1), the Breakthrough Prize Foundation and NASA (Agnostic Biosignatures award no. 80NSSC18K1140)